\documentclass[twocolumn,prl,showpacs,amsmath,amssymb]{revtex4-1}

\usepackage{graphicx}
\usepackage{bm}

\begin{document}

\preprint{APS/123-QED}

\title{Integrated liquid-core optical fibers --- ultra-efficient nonlinear liquid photonics}

\author{K. Kieu}
\email{kkieu@optics.arizona.edu}
\author{L. Schneebeli}
\author{R. A. Norwood}
\author{N. Peyghambarian}

\affiliation{
College of Optical Sciences, University of Arizona, Tucson, Arizona 85721
\\
$^*$Corresponding author: kkieu@optics.arizona.edu
}

\date{\today}

\begin{abstract}
We have developed a novel integrated platform for liquid photonics 
based on liquid core optical fiber (LCOF).  The platform is created by fusion splicing liquid core optical fiber to 
standard single-mode optical fiber making it fully integrated and practical - a major challenge that has greatly 
hindered progress in liquid-photonic applications. As an example, we report here the realization of ultralow 
threshold Raman generation using an integrated $CS_2$ filled LCOF pumped with sub-nanosecond pulses at 1064nm and 532nm. 
The measured energy threshold for the Stokes generation is $\sim$ 1nJ, about three orders of magnitude 
lower than previously reported values in the literature for hydrogen gas.
The integrated LCOF platform opens up new possibilities for 
ultralow power nonlinear optics such as efficient white light generation for displays, mid-IR generation, slow light generation, 
parametric amplification, all-optical switching and wavelength conversion using liquids that have orders of magnitude 
larger optical nonlinearities compared with silica glass.
\end{abstract}


\maketitle


The liquid state of matter is widespread in nature and possesses outstanding optical properties. 
For that reason, researchers have been attempting to harness liquids (or fluids) for different applications 
in photonics \cite{Psaltis2006,Monat2007}. However, it is quite difficult at the moment to find a real application in photonics that 
is based on this state of matter. The reason is the lack of suitable practical technology that allows working 
with long path-length liquids without the typical problems associated with diffraction, loss, and maintaining high 
intensity over the length. Liquid-core optical fiber (LCOF) was recognized very early \cite{Ippen1970,STONE1975} 
as a great platform to 
investigate and explore optical properties of liquids. LCOF is normally created by filling a small capillary with an 
appropriate liquid which should be transparent at the wavelength of interest and have a refractive index slightly higher 
than that of the capillary’s material to provide optical waveguiding via total internal reflection. Hollow-core photonics 
crystal fibers (HC-PCF) are also interesting structures where liquids or gases have been filled into the region around 
the hollow-core. Several interesting applications of HC-PCF have been recently investigated 
\cite{Benabid2002,Larsen2003,Benabid2005,Bethge2010}. 
These devices 
(LCOF and HC-PCF) make long interaction length with high optical field confinement possible, helping to enable nonlinear 
optical effects at extremely low power levels. Many interesting applications of LCOF have also been proposed and reported 
in the literature \cite{CHRAPLYVY1981,BRIDGES1982,HE1995,Dasgupta1998,Dallas2004}. These prior works demonstrated the great potential of LCOF. However, early experiments required 
free space alignment to launch the light into the core of the LCOF rendering the use of liquid unpractical. Thus, even 
though promising results were obtained, the full potential of liquid photonics were not realized due to the lack of 
best performance and it has not been used in real applications.  Here, we show that low loss fusion splicing of liquid core 
optical fiber to standard single-mode optical fiber not only results in an integrated practical and compact set-up but also 
it leads to about three-orders of magnitude better nonlinear performance compared with early experiments 
\cite{Benabid2002,Benabid2005}. As an example, 
we demonstrate ultra-low threshold of $\sim$ 1 nJ stimulated Raman scattering in an integrated $CS_2$ filled LCOF pumped with 
sub-nanosecond pulses at 1064nm and 532nm compared with 800 nJ in prior experiments \cite{Benabid2002,Benabid2005}.

There are two suitable platforms that provide long interaction length and strong field confinement namely 
capillary tubing or HC-PCF. Generally, working with a liquid filled HC-PCF or capillary requires free space 
coupling into the small guiding core. It would be desirable to be able to splice these specialty fibers to 
standard single-mode fiber while still being able to fill the liquid of interest into the hollow region. 
To date, we have tried several different approaches to achieve this. Mechanical splicing is simple 
(there is always a small gap between the fibers for liquid access) but the structure is not very stable and the 
insertion loss is normally quite high. Femtosecond laser drilling after a regular splice is another interesting approach
 but creates small debris that could block the capillary creating significant power loss. The best approach that we have 
found so far is splicing standard single-mode fiber to the capillary while leaving a small gap for liquid access (see below).

%
\begin{figure*}
\centerline{\scalebox{0.85}{\includegraphics{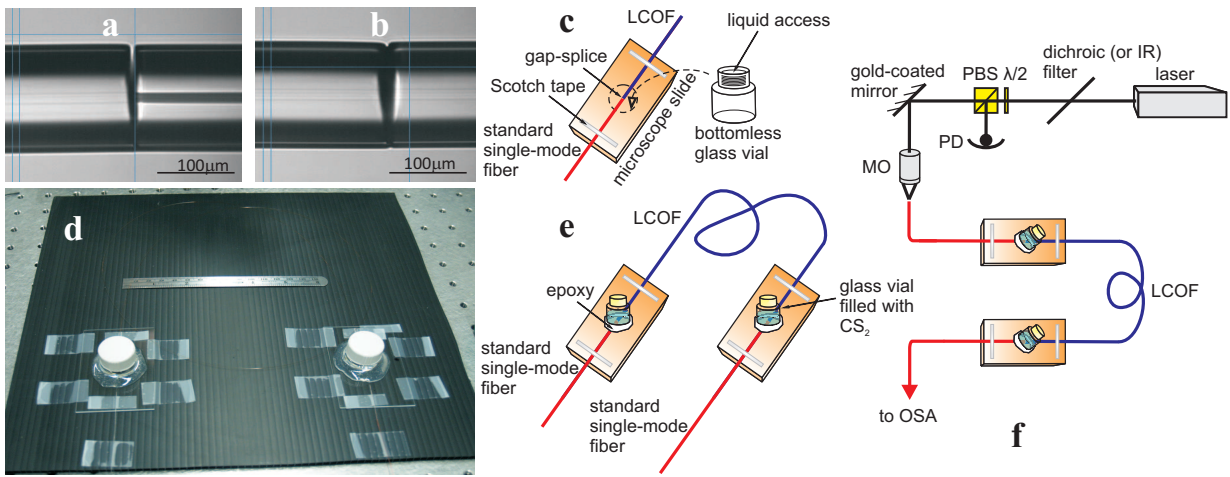}}}
\caption{(color online). Integrated LCOF preparation and stimulated Raman generation setup. a, 
Gap-splice between Corning SMF28 (left) and a 10m core LCOF (right). b, Gap-splice between two segments of Corning 
SMF28. c, Liquid access port assembly. d, Photograph of an integrated 1m long LCOF filled with $CS_2$. e, Schematic of 
an integrated LCOF filled with $CS_2$. f, Schematic diagram of the experimental setup. PBS: polarizing beam splitter; 
MO: microscope objective; PD: photodiode; OSA: optical spectrum analyzer.} 
\label{fig: schematics}
\end{figure*}

In our experiments, we use fused silica capillaries. They have different inner tube diameters 
(10 $\mu m$, 5 $\mu m$ and 2 $\mu m$), the outer diameter is around 125 $\mu m$ matching well with standard single-mode fibers. 
To perform the splice, the capillary is cleaved at zero degrees (a straight angle) but the standard fiber 
is cleaved with an angle of $\sim$ 3-10 degrees (which helps to create a small gap after the fusion splice for liquid access). 
By optimizing the parameters of the heat source hole-collapse could be avoided and a relatively strong joint with a small 
gap for liquid access could be formed [Fig.~\ref{fig: schematics}(a)]. We have made a similar splice of SMF28 to another segment 
of SMF28 
[Fig.~\ref{fig: schematics}(b)] and the loss of the transition was measured to be only around 0.1 dB. This indicates that the small gap left 
after the splice does not create appreciable loss. Due to the fact that the fibers are fused only in a small area the joint 
is not as strong as a standard fusion splice. However, the structure could be handled and transported (with some care) to a 
holder (access port) where it can be fixed permanently using an appropriate epoxy [Fig.~\ref{fig: schematics}(c)]. A similar gap-splice is made 
at the other end of the capillary tubing at a certain length to create the second access port for liquid filling. After 
filling with the liquid the access ports could be sealed and the structure remains stable for weeks 
[Figs.~\ref{fig: schematics}(d)-\ref{fig: schematics}(e)]. 
This technique integrates LCOF with standard optical fiber in a compact 
and stable package without the need for free space alignment or complicated liquid handling procedures. We have successfully 
used these devices in a few applications including Raman spectroscopy, supercontinuum generation and low threshold Raman 
generation, as described below.

Stimulated Raman scattering (SRS) is a nonlinear optical process in which an incoming photon interacts with a 
coherently excited system resulting in generation of new frequencies with very high conversion efficiency. 
SRS is useful in wavelength conversion, amplification, generation of ultra-short (atto-second) optical pulses 
\cite{Harris1998,Sokolov1999,Rong2005,Couny2007,Jones2011}, 
precision spectroscopy and microscopy \cite{Dudovich2002,Cheng2004,Freudiger2008}. SRS occurs when the pump power reaches a certain threshold to provide enough 
gain for the Stokes wave. The threshold peak power can be estimated using the following simple formula \cite{AgrawalBook} 
\begin{equation}
P^{\mathrm{thr}} = 16 A_\mathrm{eff}/(g_R L_\mathrm{eff})  , 
\label{eq: threshold estimate}
\end{equation}
where in the present case the effective area is $A_\mathrm{eff} = 3.14 \cdot 10^{-12} \, \mathrm{m^2}$, the Raman gain 
coefficient for $CS_2$ is
$g_R=1.25 \cdot 10^{-10} \, \mathrm{m/W}$ at 532 nm \cite{Colles1969}, and the 
effective interaction length is $L_\mathrm{eff} = 1 \, \mathrm{m}$. 

We used a frequency doubled microchip laser (JDSU) that emits sub-nanosecond ($\sim$ 500ps) Q-switched pulses as the pump source. 
The repetition rate of the source is $\sim$ 1.5kHz. The laser beam is coupled into a short piece (1m) of Corning HI1060 fiber 
using a microscope objective and translation stages as shown in Fig.~\ref{fig: schematics}(f). The coupling efficiency 
is $\sim$ $50 \%$. 
The coupled power into the standard fiber is adjusted using a combination of a half waveplate and a polarizing beam splitter. 
For experiments in the visible, an IR filter is used to remove the residual 1064nm pulses. For experiments using 1064nm 
wavelength, the IR filter is replaced by a dichroic filter to remove the visible green beam.

\begin{figure*}[ht]
\centerline{\scalebox{0.8}{\includegraphics{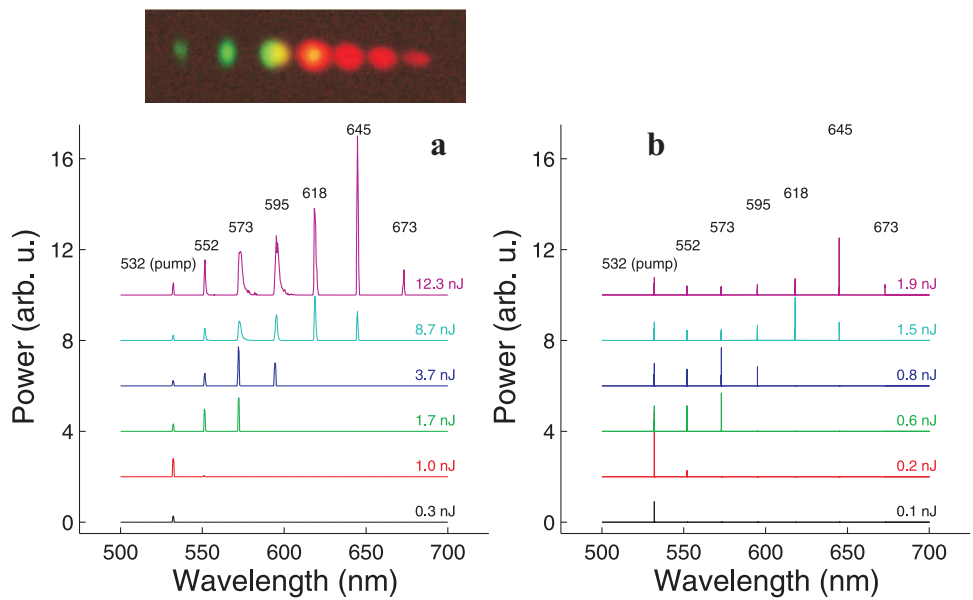}}}
\caption{(color online). Experimental and simulation results for 532nm pumping. 
Measured a) and calculated b) evolution of the output spectrum as the pump pulse energy is increased, 
for a pump wavelength of 532 nm. Inset of a): Photograph of the Raman lines separated in space using a prism.} 
\label{fig: waterfall 532 nm}
\end{figure*}

After the coupling step, light is launched into the integrated LCOF by simply splicing to the HI1060 fiber at one end. 
Figure~\ref{fig: waterfall 532 nm}(a) shows the evolution of the laser output after propagating through the $CS_2$ filled LCOF. 
Several orders of Stokes 
Raman generation are seen as the pump power is increased. The threshold of stimulated Raman generation is at an extremely 
low pump pulse energy of $\sim$ 1nJ. This energy threshold is near three orders of magnitude lower than the 
value reported for hydrogen filled HC-PCF of similar length \cite{Benabid2002}. The ultralow threshold observed in our experiment can be 
attributed to the unprecedented confinement of the optical field ($\sim$ 2micron core size), the long propagation length ($\sim$ 1m), 
and the large Raman gain coefficient of $CS_2$.
Using Eq.~(\ref{eq: threshold estimate}), we obtain an estimated threshold pump peak power of 0.4W at 532 nm pump wavelength which is 
consistent with the low-power threshold observed experimentally (2~W). We obtain the same threshold value when we 
numerically solve the coupled amplitude equations in the picosecond regime \cite{AgrawalBook}.

In Fig.~\ref{fig: waterfall 532 nm}(a), up to 6 orders of stimulated 
Raman scattering are generated at only $\sim$ 12nJ of coupled pump pulse energy. 
We expect to see more Raman lines at higher pulse energies but they would be out of the measurement 
range of the OSA. We also observed broadening of the higher order Stokes lines. 
Broadening of stimulated Raman lines was also observed and explained by N. Bloembergen in Ref.~\cite{Bloembergen1966}. 
Detailed understanding of the 
difference in the level of broadening for different orders is quite complex and is beyond the scope of this paper.

In the inset of Fig.~\ref{fig: waterfall 532 nm}(a), we nicely see multiple Raman 
lines that are separated in space using a prism made of SF10 glass. Due to the strong residual absorption of $CS_2$ at 532nm 
we observed gradual reduction in transmission of the pump light due to heating. The capillary was eventually 
(in matter of $\sim$ 30 minutes) blocked and re-splicing of the fibers was required to observe the stimulated Raman 
generation process again. This thermally induced degradation was not observed when the fiber was pumped at 1064nm 
even at the highest available pulse energy ($\sim$ 300nJ) for a few weeks.

Calculations of the higher-order Stokes generation are carried out based on the coupled equations presented in 
Refs.~\cite{Bertoni1997,Min2003}. 
Explicitly, these coupled equations read

\begin{align}
\frac{\partial A_p}{\partial z} &= -\frac{g_p}{2} |A_{s,1}|^2 A_p ,
\\\notag
\frac{\partial A_{s,l}}{\partial z} &= \frac{g_{s,l}}{2} |A_{s,l-1}|^2 A_{s,l} - \frac{g_{s,l}}{2} |A_{s,l+1}|^2 A_{s,l}
\\
&(l=1,\ldots,M-1; A_{s,0} \equiv A_p) ,
\\
\frac{\partial A_{s,M}}{\partial z} &= \frac{g_{s,M}}{2} |A_{s,M-1}|^2 A_{s,M}.
\end{align}
The wavelength-dependent normalized [with respect to the effective area $g = g_R / A_\mathrm{eff}$]
Raman gain coefficients $g_p$ and $g_{s,l}$  
for the space-time dependent pump 
$A_p(z,t)$ and each Stokes line $A_{s,l}(z,t)$ up to the Mth order enter, respectively, 
while dispersion and nonlinearities have been neglected as a good approximation due to the long pump duration. 
Figure~\ref{fig: waterfall 532 nm}(b) shows the calculated output spectra for increasing pump pulse energies.  
We notice that the number and position of Raman lines and 
their relative heights are well reproduced by theory.
We observe that up to 6 Raman orders are visible at a pump pulse energy of $\sim$ 2 nJ. 
At lower pulse energies, only lower-order Raman lines are present, in good agreement with the experiment 
[Fig.~\ref{fig: waterfall 532 nm}(a)]. 
The difference between experiment and theory can be explained by the variation of the core size along the fiber, 
uncertainties in the Raman gain coefficient and the partially polarized input pump (due to propagation through a long 
segment of non-polarization preserving fiber).

%
\begin{figure}
\centerline{\scalebox{0.65}{\includegraphics{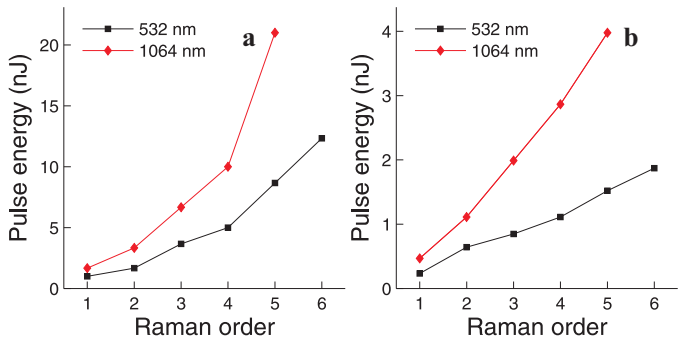}}}
\caption{(color online). Pulse energy threshold for generation of different Raman orders. 
The measured a) and calculated b) pulse energies corresponding to generation of different orders of Raman generation.} 
\label{fig: Raman order}
\end{figure}

The measured energy corresponding to the generation of different orders of stimulated Raman scattering is shown in 
Fig.~\ref{fig: Raman order}(a). The threshold energy for the 1064 nm pump wavelength is higher than in the case of the 532nm pump, 
which is expected since the Raman gain coefficient is lower at longer wavelength \cite{AgrawalBook,YablonBook}. 
Figure~\ref{fig: Raman order}(b) 
shows the corresponding 
calculations which are in good agreement with Fig.~\ref{fig: Raman order}(a). We observe a linear dependence in the theory as 
observed in 
experiment for low pump powers [Fig.~\ref{fig: Raman order}(a)]. The deviation from the linear dependence as seen in the 
experimental data [Fig.~\ref{fig: Raman order}(a)]
at higher pump powers is conjectured to come from the higher absorption loss (due to water contamination) at longer wavelengths, 
or possibly reduced confinement and hence increased effective area for the longer wavelength Raman lines.

The reduction of Raman threshold power in LCOF should allow the use of low cost diode pumped lasers to create new frequencies. 
For example, we can envision the use of this platform to generate the required laser lines (red, green and blue) for future 
color TV display. Atto-second pulse generation via stimulated Raman scattering should also benefit from the low power 
threshold demonstrated here as well. The development of integrated LCOF is likely to lead to rapid progress of other 
important applications in nonlinear optics such as supercontinuum generation, mid-IR generation, slow light, all-optical 
switching and wavelength conversion. The prospects from practical liquid photonics appear to be quite bright. We further 
note that the recent development of organic materials \cite{Hales2010} with tremendous molecular hyper-polarizability (1000x$CS_2$) provides 
a route to pJ level operating energies. In addition, the gap-splicing technique reported here can be used with HC-PCF to 
open up the possibility of changing the gas under investigation on the same platform. This should be useful in applications 
involving gas sensing or breath analysis \cite{Thorpe2006}.

Acknowledgement: This work was supported by the DARPA ZOE program (Grant No. W31P4Q-09-1-0012), 
the CIAN ERC (Grant No. EEC-0812072), and the USAF, AFRL COMAS MURI program (Grant No. FA9550-10-1-0558).


%

\end{document}